\documentclass[tightenlines,a4paper,altaffilletter,twoside,secnumarabic,
               showpacs,twocolumn,floatfix,nofootinbib,prd]{revtex4}

\usepackage[english]{babel}
\usepackage{amsmath}
\usepackage{amssymb}
\usepackage{epsfig}
\usepackage{graphicx}

\newcommand{\bra}[1]{\ensuremath{\langle #1 |}}   
\newcommand{\ket}[1]{\ensuremath{| #1 \rangle}}   

\newcommand{\emu}{\ensuremath{\nu_e \rightarrow \nu_\mu}}
\newcommand{\etau}{\ensuremath{\nu_e \rightarrow \nu_\tau}}
\newcommand{\mue}{\ensuremath{\nu_\mu \rightarrow \nu_e}}
\newcommand{\mumu}{\ensuremath{\nu_\mu \rightarrow \nu_\mu}}
\newcommand{\mutau}{\ensuremath{\nu_\mu \rightarrow \nu_\tau}}

\newcommand{\ldm}{\ensuremath{\Delta m_{31}^2}}
\newcommand{\sdm}{\ensuremath{\Delta m_{21}^2}}

\begin{document}

\title{Discovery reach for non-standard interactions in a neutrino factory}
\author{Joachim Kopp}    \email[Email: ]{jkopp@mpi-hd.mpg.de}
\author{Manfred Lindner} \email[Email: ]{lindner@mpi-hd.mpg.de}
\author{Toshihiko Ota}   \email[Email: ]{toshi@mpi-hd.mpg.de}
\affiliation{Max--Planck--Institut f\"ur Kernphysik, \\
             Postfach 10 39 80, 69029 Heidelberg, Germany}

\begin{abstract}
 We study the discovery reach for Non-Standard Interactions (NSIs)
 in a neutrino factory experiment. After giving a theoretical, but
 model-independent, overview of the most relevant classes of NSIs,
 we present detailed numerical results for some of them. Our
 simulations take into account matter effects, uncertainties in
 the neutrino oscillation parameters, systematical errors, parameter
 correlations, and degeneracies. We perform scans of the parameter space,
 and show that a neutrino factory has excellent prospects of detecting
 NSIs originating from new physics at around 1~TeV, which is a scale
 favored by many extensions of the standard model. It will also turn
 out that the discovery reach depends strongly on the standard and
 non-standard CP violating phases in the Lagrangian.
\end{abstract}

\maketitle

\section{Introduction}

Huge efforts are currently undertaken to design new long-baseline neutrino
experiments to precisely measure the three-flavor oscillation parameters, in
particular the yet unknown mixing angle $\theta_{13}$, the CP violating phase
$\delta_{\rm CP}$, and the sign of the atmospheric mass squared difference
$\ldm$. However, the excellent accuracy with which the planned setups can
measure the oscillation probabilities, will also allow for the detection of
new sub-leading effects, such as mixing with sterile neutrinos, a non-unitary
Pontecorvo-Maki-Nakagawa-Sakata (PMNS) matrix, neutrino decay, the decoherence
effect, CPT violation, or mass-varying neutrinos. Furthermore, many extensions
of the standard model predict new, effective four-Fermi interactions involving
neutrinos, on which we will focus in this paper. General phenomenological studies
of these non-standard interactions (NSIs) have been conducted in~\cite{Wolfenstein:1977ue,
Valle:1987gv,Guzzo:1991hi,Roulet:1991sm,Bergmann:1999rz,Hattori:2002uw,
Garbutt:2003ih,Blennow:2005qj}, and specific models are discussed
in~\cite{DeGouvea:2001mz,Ota:2005et}. After a work by
Grossman~\cite{Grossman:1995wx}, which pointed out the importance of NSIs for
neutrino oscillation experiments, many authors have investigated their impact
in the context of solar neutrinos~\cite{Bergmann:2000gp,Berezhiani:2001rt,Friedland:2004pp,
Miranda:2004nb}, atmospheric neutrinos~\cite{Gonzalez-Garcia:1998hj,Bergmann:1999pk,Fornengo:2001pm,
Gonzalez-Garcia:2004wg,Friedland:2004ah,Friedland:2005vy}, conventional and
upgraded neutrino beams~\cite{Bergmann:1998ft,Ota:2001pw,Ota:2002na,Honda:2006gv,
Kitazawa:2006iq,Friedland:2006pi,Blennow:2007pu}, neutrino
factories~\cite{Ota:2001pw,Gonzalez-Garcia:2001mp,Huber:2001zw,Gago:2001xg,
Huber:2002bi,Campanelli:2002cc,Blennow:2005qj,Bueno:2000jy}, beta beams~\cite{Adhikari:2006uj},
supernova neutrinos~\cite{Fogli:2002xj,Duan:2006jv}, cosmological relic
neutrinos~\cite{Mangano:2006ar}, $e^+ e^-$ colliders~\cite{Berezhiani:2001rs},
neutrino-electron scattering~\cite{Barranco:2005ps}, and neutrino-nucleus
scattering~\cite{Barranco:2005yy,Barranco:2007tz}. Existing experimental bounds
are presented in~\cite{Davidson:2003ha}.

In this article, we will discuss in particular the discovery potential
of a neutrino factory~\cite{Geer:1997iz,Albright:2000xi,Apollonio:2002en,
Huber:2006wb,Huber:2001de,Freund:2001ui,Freund:2000ti}, which is currently
the most advanced technology discussed in neutrino physics, and would have
a precision of $\mathcal{O} (1\text{--}0.1) $\% on some of the oscillation
probabilities. We will show that this translates into a sensitivity to
NSIs originating from scales of up to several TeV. We will focus only on
non-standard interactions (NSIs) which violate lepton flavor. Existence
of such interactions would typically induce not only effects in the neutrino
sector, but also charged lepton flavor violating processes like
$\mu \rightarrow 3e$. However, in the charged lepton sector, the signal is
proportional to the \emph{square} of the non-standard coupling, while in an
oscillation experiment, interference between the standard and non-standard
amplitudes will also induce terms which are linear in the coupling constant
and can therefore be expected to be easier to detect.

A long baseline neutrino oscillation experiment consists of three stages:
beam production (source), beam propagation through the Earth, and neutrino
detection at the far site. Here, we are going to consider NSIs which modify
only one of these aspects at a time. In principle, there may also exist
combined effects of several new processes, but these will be suppressed
by higher powers of the small coupling constants.

The paper is organized as follows: In Sec.~\ref{Sec:Analytic}, we will
describe the NSIs analytically, but in a model-independent way.
Afterwards, in Sec.~\ref{Sec:Numerical}, we will present detailed
numerical results on non-standard modifications to the neutrino production
and propagation amplitudes. We have performed sophisticated simulations
with a modified version of the GLoBES software~\cite{Huber:2004ka,Huber:2007ji},
taking into account systematical errors and correlations
between all standard and non-standard oscillation parameters. Some of
these correlations will turn out to be very strong, so our final results on
the NSI discovery reach of a neutrino factory will strongly depend on
the true parameter values. We will summarize our results in
Sec.~\ref{Sec:Conclusion} and draw some conclusions.

\section{Non-standard interactions in neutrino oscillations}
\label{Sec:Analytic}

\begin{figure*}
  \begin{center}
    \includegraphics{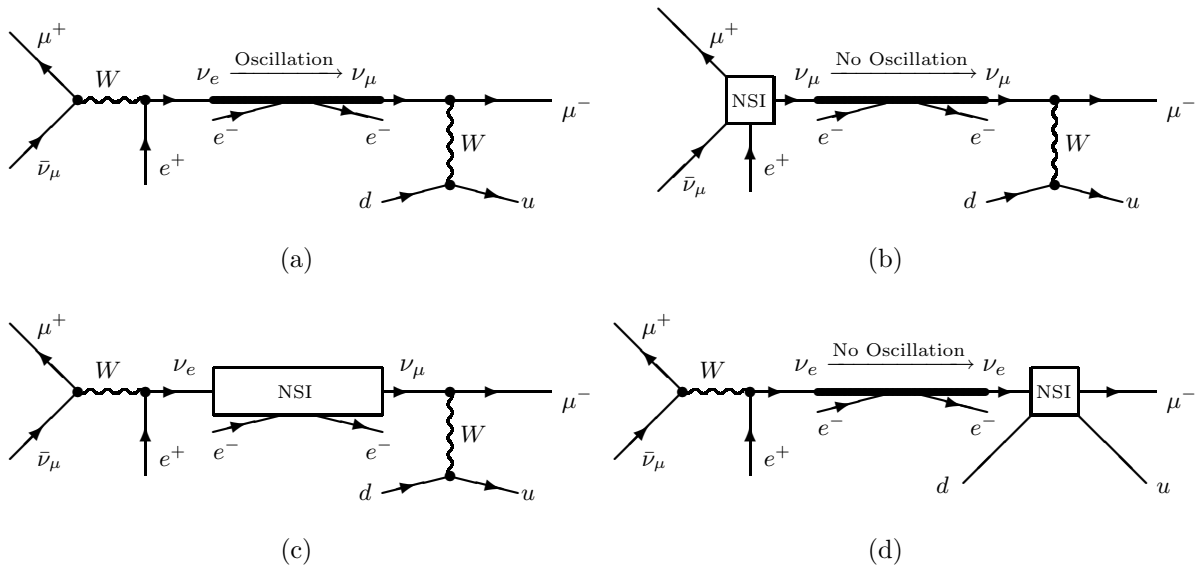}
  \end{center}
  \caption{(a): The golden channel oscillation process in a neutrino factory
    (cf.\ Eq.~\eqref{eq:Amp-golden-SO}); (b) -- (d): Non-standard contributions
    to the golden channel (cf.\ Eqs.~\eqref{eq:Amp-golden-s}
    to~\eqref{eq:Amp-golden-d}).}
  \label{Fig:diagrams}
\end{figure*}

In the context of neutrino factory experiments, one usually considers
the ``golden oscillation channel'' $\emu$, the ``silver channel'' $\etau$,
the ``platinum channel'' $\mue$, and the disappearance channel $\mumu$
(see e.g.~\cite{Huber:2006wb} and references therein).
Of these, the golden channel is most important for the discovery of small
effects such as $\theta_{13}$-oscillations, CP violation, but also
non-standard interactions, because it is an appearance channel, and
because it is technically more easily accessible than the silver and
platinum channels. Therefore, we will focus on the golden channel and consider
only NSIs influencing the corresponding process chain
\begin{align}
  \mu^{+} \rightarrow \nu_{e} \xrightarrow[]{\text{Osc.}}
   \nu_{\mu} \rightarrow \mu^{-}.
  \label{eq:Amp-golden-SO}
\end{align}
NSIs can modify the production, oscillation, and detection of neutrinos,
so that the following alternative processes to Eq.~\eqref{eq:Amp-golden-SO}
can occur:
\begin{align}
&
 \mu^{+} \xrightarrow[]{\text{NSI}}
 \nu_{\mu} \xrightarrow[]{\text{No osc.}}
 \nu_{\mu} \xrightarrow[]{}
 \mu^{-},
\label{eq:Amp-golden-s} \\
&
 \mu^{+} \xrightarrow[]{}
 \nu_{e} \xrightarrow[\text{NSI}]{\text{No osc.}}
 \nu_{\mu} \xrightarrow[]{}
 \mu^{-},
\label{eq:Amp-golden-memu} \\
&
 \mu^{+} \xrightarrow[]{}
 \nu_{e} \xrightarrow[]{\text{No osc.}}
 \nu_{e} \xrightarrow[]{\text{NSI}}
 \mu^{-}.
\label{eq:Amp-golden-d}
\end{align}
These processes are illustrated diagrammatically in Fig.~\ref{Fig:diagrams}.

Since the initial and final states are the same in Eq.~\eqref{eq:Amp-golden-SO}
and Eqs.~\eqref{eq:Amp-golden-s} to~\eqref{eq:Amp-golden-d}, interference
can occur on the level of the amplitudes~\cite{Grossman:1995wx,Wolfenstein:1977ue,
Valle:1987gv,Guzzo:1991hi,Roulet:1991sm,Bergmann:1999rz,Gago:2001xg}.
This will enhance the magnitude of the new effects compared to scenarios
where the NSIs are added non-coherently.

Non-standard interactions are typically generated by new physics at very high energy
scales, so for a neutrino factory operating in the low-energy regime, they
can be expressed as model-independent four-Fermi interactions:
\begin{align}
  \mathcal{L}_{\text{NSI}} \ = & \ 
    \frac{G_{F}}{\sqrt{2}} \, \epsilon^{s\mp}_{e\mu}
      \left\{ \bar{\nu}_\mu \gamma^{\rho} (1 - \gamma^{5}) \nu_{\mu} \right\}
      \left\{ \bar{\mu} \gamma_{\rho} (1 \mp \gamma^{5}) e \right\}
    \nonumber \\
  & \hspace{-0.8 cm}
    + \sum_{f = e,u,d} \frac{G_{F}}{\sqrt{2}} \, \epsilon^{m,f\mp}_{e\mu}
      \left\{ \bar{\nu}_e \gamma^{\rho} (1 - \gamma^{5}) \nu_{\mu} \right\}
      \left\{ \bar{f} \gamma_{\rho} (1\mp\gamma^{5}) f \right\}
    \nonumber \\
  & \hspace{-0.8 cm}
    + \frac{G_{F}}{\sqrt{2}} \, \epsilon^{d\mp}_{e\mu}
        \left\{ \bar{\mu} \gamma^{\rho} (1 - \gamma^{5}) \nu_{e} \right\}
        \left\{ \bar{u} \gamma_{\rho} (1\mp\gamma^{5}) d \right\} + {\rm h.c.}
    \label{eq:Lag-LFV}
\end{align}
Here, $G_F$ denotes the Fermi constant, $\nu_e$ and $\nu_\mu$
are the neutrino fields, $e$ and $\mu$ are the charged lepton
fields, and $u$, $d$ are quark fields. Finally, the magnitude of the NSIs
is parameterized by $\epsilon^{s\mp}_{e\mu}$ for effects in the neutrino
source, by $\epsilon^{m,f\mp}_{e\mu}$ for non-standard matter effects on the
oscillation, and by $\epsilon^{d\mp}_{e\mu}$ for modifications
to the detection process. In contrast to a previous study~\cite{Gago:2001xg},
we allow the $\epsilon$ parameters to be complex. Then, Eq.~\eqref{eq:Lag-LFV}
implies that the $\epsilon$ parameters for anti-neutrino processes are the
complex conjugates of those for neutrino processes.

The expected magnitude of the $\epsilon$ parameters can be estimated as
follows~\cite{Gonzalez-Garcia:2001mp}: If we assume the non-standard interactions
to arise at a scale $M_{\rm NSI}$, the effective vertices in Eq.~\eqref{eq:Lag-LFV}
will be suppressed by $1 / M_{\rm NSI}^2$ in the same way as the standard weak
interactions are suppressed by $1 / M_{\rm W}^2$. Therefore we expect
\begin{align}
  |\epsilon| \sim \frac{M_{\rm W}^2}{M_{\rm NSI}^2}.
  \label{eq:epsilon-estimate}
\end{align}

The NSIs in the beam source, given by the first line of Eq.~\eqref{eq:Lag-LFV},
effectively promote the initial neutrino state from a pure flavor eigenstate into
the mixed state
\begin{align}
  \ket{\nu_e^{(s)}} = \ket{\nu_{e}} + \left\{
        \epsilon^{s -}_{e\mu} + \epsilon^{s +}_{e\mu}
        \times \mathcal{O}\left(\frac{m_{\mu} m_{e}}{E_{\mu}E_{e}} \right)
       \right\} \ket{\nu_{\mu}}.
  \label{eq:mod-i-state}
\end{align}
Similarly, the detector will project out the mixed state
\begin{align}
 \bra{\nu^{(d)}_{\mu}} = \bra{\nu_{\mu}}  + \left\{
    \epsilon^{d -}_{e\mu} + \epsilon^{d +}_{e\mu} \times
    \mathcal{O}\left(\frac{m_{u} m_{d}}{E_{u} E_{d}} \right)
   \right\} \bra{\nu_{e}}
 \label{eq:mod-f-state}
\end{align}
rather than the pure flavor eigenstate $\bra{\nu_\mu}$.
The kinematics of a neutrino factory experiment is such that
in most regions of the phase space, $E_{e,\mu,u,d} \gg m_{e,\mu,u,d}$.
This remains true even if we take into account that $E_e$ and $E_\mu$ as
well as $E_u$ and $E_d$ are not independent, and if we allow $E_u$
and $E_d$ to be slightly off-shell due to QCD effects in the nucleus.
We do not consider these hadronic effects here, but we remark that they
will generally be soft compared to the primary neutrino interaction.
Since interference of standard and non-standard processes can only
occur if \emph{all} initial and final state particles have the same
chirality, it follows that the $(V-A)(V+A)$ type NSIs are suppressed
by the helicity factors
$\mathcal{O}\left(m_{\mu} m_{e} / E_{\mu}E_{e} \right)$ resp.\
$\mathcal{O}\left(m_{u} m_{d} / E_{u} E_{d} \right)$. Therefore, we will take
\begin{align}
  \epsilon^{s}_{e\mu} &= \epsilon^{s -}_{e\mu},  \\
  \epsilon^{d}_{e\mu} &= \epsilon^{d -}_{e\mu}.
\end{align}
in the following.

Note that the $\epsilon$ parameters do not necessarily form unitary matrices,
so that the source and detection states, in general, do not form complete
sets of basis vectors in the Hilbert space:
\begin{align}
  \sum_{\alpha = e,\mu,\tau} \ket{\nu^{(s)}_{\alpha}} \bra{\nu^{(s)}_{\alpha}}
          \neq 1, \qquad
  \sum_{\alpha = e,\mu,\tau} \ket{\nu^{(d)}_{\alpha}} \bra{\nu^{(d)}_{\alpha}}
          \neq 1.
\end{align} 
However, we do require the PMNS matrix to be unitary, so that the standard
mass and flavor eigenstates at least form a basis of the subspace of states
participating in oscillations. Thus, the neutrino propagation does not violate
unitarity, while the production and detection processes may do so. With these
assumptions, neutrino oscillations can be described as usual by a hermitian
$3 \times 3$ Hamiltonian~\cite{Gago:2001xg}, that contains, however, an extra
term $H_{\rm NSI}$ due to the second line of Eq.~\eqref{eq:Lag-LFV}. Thus we
can write
\begin{align}
  H          &= H_{\rm SO} + |\epsilon^{m}_{e\mu}| H_{\rm NSI}(\epsilon^{m}_{e\mu}),
  \label{eq:H_emu}  \\
  H_{\rm SO} &\equiv U_{\alpha i} \!
                     \begin{pmatrix}
                       0 &                 & \\
                         & \frac{\sdm}{2E} & \\
                         &                 & \frac{\ldm}{2E}
                     \end{pmatrix} \!
                     (U^{\dagger})_{i \beta}
                   + \begin{pmatrix}
                       \frac{a}{2E} &   &          \\
                                    & \!\! 0 &     \\
                                    &        & \!0
                     \end{pmatrix},         \\
  H_{\rm NSI}(\epsilon^{m}_{e\mu}) &\equiv \frac{a}{2E}
                  \begin{pmatrix}
                    0 & {\rm e}^{{\rm i} \arg[\epsilon^{m}_{e\mu}]} & \\
                    {\rm e}^{-{\rm i}\arg[\epsilon^{m}_{e\mu}]} & 0 &\\
                      & & 0
                  \end{pmatrix}.
  \label{eq:H_emu_NSI}
\end{align}
Here, $H_{\rm SO}$ contains the standard oscillations (SO), and $a$ is the effective
matter potential, which we assume to be constant in the following. The effective
NSI coupling $\epsilon^{m}_{e\mu}$ is related to the parameters
$\epsilon^{m \mp, f}_{e\mu}$ from Eq.~\eqref{eq:Lag-LFV} by the formula
\begin{multline}
  \epsilon^{m}_{e\mu} = (\epsilon^{m,e+}_{e\mu} + 3\epsilon^{m,u+}_{e\mu} +
                                                  3\epsilon^{m,d+}_{e\mu}) + \\
                        (\epsilon^{m,e-}_{e\mu} + 3\epsilon^{m,u-}_{e\mu} +
                                                  3\epsilon^{m,d-}_{e\mu}).
\end{multline}
This relation can be understood if we assume the numbers of protons, neutrons,
and electrons to be the same in the Earth matter and consider only the
effect of valence quarks. Furthermore, we have made use of the fact that the
spin and momentum average of the Earth is zero, so that only the components of
the vector interactions in the second line of Eq.~\eqref{eq:Lag-LFV} are relevant,
and contribute equally.

In principle, $H_{\rm NSI}$ can also contain other
non-zero entries besides $\epsilon^{m}_{e\mu}$. In combination
with standard oscillations, these can lead to process chains like
\begin{align}
 \mu^{+} \xrightarrow[]{}
 \nu_{e} \xrightarrow[\text{NSI}]{\text{No osc.}}
 \nu_{\tau} \xrightarrow[]{\text{Osc.}}
 \nu_{\mu} \xrightarrow[]{}
 \mu^{-}.
 \label{eq:Amp-golden-metau}
\end{align}
In the remainder of this section, we will neglect such contributions for
conciseness, but we will exemplarily consider effects proportional
to $\epsilon^{m}_{e\tau}$ in our numerical analysis in Sec.~\ref{Sec:epsilon-m-etau}.
A systematic study of non-standard Hamiltonians like
Eq.~\eqref{eq:H_emu_NSI} is given in~\cite{Blennow:2005qj}.

The amplitude of the flavor transition $\nu_{\alpha} \rightarrow \nu_{\beta}$
can be calculated from the propagation Hamiltonian Eq.~\eqref{eq:H_emu}
by\footnote{Here we regard the NSI parameters $\epsilon^{s,m,d}_{e\mu}$ to
be small perturbations. In~\cite{Kitazawa:2006iq} the authors pointed out that, from the
current experimental limits, the NSIs might even dominate over the oscillation
effect in a $\mutau$ oscillation experiment. In such a situation, this
perturbative expansion would no longer be valid.}
\begin{align}
  \bra{\nu_{\beta}} {\rm e}^{-{\rm i} H L} \ket{\nu_{\alpha}} =
      (S_{\rm SO})_{\beta \alpha}  + |\epsilon^{m}_{e\mu}| 
      \{ S_{\rm NSI}(\epsilon^{m}_{e\mu}) \}_{\beta \alpha}
    + \mathcal{O}(\epsilon^{2}).
  \label{eq:NSI-S-Matrix}
\end{align}
Here, the standard oscillation amplitude $S_{\rm SO}$ reads
\begin{align}
  S_{\rm SO} \equiv {\rm e}^{-{\rm i} H_{\rm SO} L},
\end{align}
and the amplitude induced by the non-standard matter effects, $S_{\rm NSI}$,
is given by the perturbative expansion
\begin{multline}
 \{S_{\rm NSI} (\epsilon^{m}_{e\mu})\}_{\beta \alpha} = 
   \sum_{\gamma=e,\mu,\tau} \!\! (S_{\rm SO})_{\beta \gamma} \\
     \cdot \left\{
       - {\rm i} \int_{0}^{L} {\rm d}x \, {\rm e}^{{\rm i} H_{\rm SO} x } 
              H_{\rm NSI} (\epsilon^{m}_{e\mu}) {\rm e}^{-{\rm i} H_{\rm SO} x }
   \right\}_{\gamma \alpha}.
  \label{eq:S_NSI}
\end{multline}
In our case, the $(e, \mu)$ and $(\mu, e)$ elements of 
$\{H_{\rm NSI}(\epsilon^{m}_{e\mu}) \}_{\beta \alpha}$ are non-zero,
so the golden-channel flavor transition $\nu_{e} \rightarrow \nu_{\mu}$
can occur even in the absence of standard oscillations.

If, as a last step, we replace the initial and final states in
Eq.~\eqref{eq:NSI-S-Matrix} by the modified states from Eqs.~\eqref{eq:mod-i-state}
and~\eqref{eq:mod-f-state}, we obtain the transition probability up to
first order in the $\epsilon$ parameters as
\begin{align}
  P(\nu_{e}^{(s)} \rightarrow \nu_{\mu}^{(d)}) &=
    \left| \bra{\nu^{(d)}_{\mu}} {\rm e}^{- {\rm i} H L} \ket{\nu^{(s)}_{e}} \right|^{2}
                                                                           \\
  &= |(S_{\rm SO})_{\mu e}|^{2} \nonumber \\
  & \hspace{0.5 cm} +
     2 |\epsilon^{m}_{e\mu}| {\rm Re}
      \left[ (S_{\rm SO})_{\mu e}^{*}
            \{S_{\rm NSI}(\epsilon^{m}_{e\mu})\}_{\mu e} \right] \nonumber \\
  & \hspace{0.5 cm} +  2 |\epsilon^{s}_{e\mu}| {\rm Re}
      \left[ (S_{\rm SO})_{\mu e}^{*} (S_{\rm SO})_{\mu\mu}
             {\rm e}^{{\rm i} \arg[\epsilon^{s}_{e\mu}]} \right] \nonumber \\
  & \hspace{0.5 cm} +  2 |\epsilon^{d}_{e\mu}| {\rm Re}
      \left[ (S_{\rm SO})_{\mu e}^{*} (S_{\rm SO})_{ee}
             {\rm e}^{{\rm i} \arg[\epsilon^{d}_{e\mu}]} \right] \nonumber \\
  & \hspace{0.5 cm} +  \mathcal{O}(\epsilon^{2}).
\label{eq:OscProb-NSI}
\end{align}
The zeroth order term represents the standard oscillation probability, while the
first order terms contain the contributions from the different types of NSIs.

In this work, we are interested in the discovery potential for non-standard
effects, i.e.\ in the prospects of identifying the tiny NSI contribution on the large
standard oscillation background.
If only the terms proportional to $\epsilon^{m}_{e\mu}$ are present, this can be
achieved by exploiting the different spectral structure of the signal and background
events~\cite{Ota:2001pw}.
If we expand the oscillation amplitudes up to first order in $1/E$, we find that
$(S_{\rm SO})_{\mu e} \sim 1/E$, while according to Eq.~\eqref{eq:S_NSI},
$\{S_{\rm NSI} (\epsilon^{m}_{e\mu})\}_{\mu e} \sim (S_{\rm SO})_{\mu \mu} \sim 1$.
Hence, the first (standard) term in Eq.~\eqref{eq:OscProb-NSI} behaves as $1/E^2$,
while the second (non-standard) term is proportional to $1/E$.
The situation is quite different for effects proportional to $\epsilon^{m}_{e\tau}$,
since for these, the non-standard terms $\{S_{\rm NSI} (\epsilon^{m}_{e\tau})\}_{\mu e}$
will contain a factor $(S_{\rm SO})_{\mu \tau} \sim 1/E$, so their lowest order
energy dependence is identical to that of the standard oscillations. Therefore,
the discovery reach of a neutrino factory for $\epsilon^{m}_{e\tau}$ will be
worse than that for $\epsilon^{m}_{e\mu}$.

For non-standard effects parameterized by $\epsilon^{s}_{e\mu}$ and
$\epsilon^{d}_{e\mu}$, we can read off from Eq.~\eqref{eq:OscProb-NSI} that
we are again in a favorable situation, since
$(S_{\rm SO})_{\mu e}^{*} (S_{\rm SO})_{\mu\mu} \sim 1/E$ and
$(S_{\rm SO})_{\mu e}^{*} (S_{\rm SO})_{ee} \sim 1/E$.

\section{Detecting non-standard interactions in a neutrino factory}
\label{Sec:Numerical}

To obtain reliable estimates for the prospects of discovering non-standard
interactions in a neutrino factory, we have performed detailed numerical
simulations with a modified version of the {\sf GLoBES}
software~\cite{Huber:2004ka,Huber:2007ji}. We use a neutrino factory setup
based on {\sf NuFact2} from~\cite{Huber:2002mx}, with a parent muon energy
of 50~GeV and a baseline of 3000~km. The total running time is 8~years
(4~years in the neutrino mode, 4~years in the anti-neutrino mode), and
the number of stored muons per year is $1.066 \cdot 10^{21}$. The detector is 
a 50~kt magnetized iron calorimeter, and the cross sections are based
on~\cite{Messier:1999kj,Paschos:2001np}. Both the wrong-sign muon
appearance channel (``golden channel'') and the muon disappearance channel
are taken into account. We have incorporated the backgrounds due to
neutral current events and muon charge misidentification.

We quantify the performance of an experiment by introducing the \emph{discovery
reach} for non-standard interactions, which is defined as the minimal
magnitude of the $\epsilon$ parameters, for which the expected experimental data
is no longer consistent with a standard oscillation fit.

Following the statistical procedure described in the appendix
of~\cite{Huber:2002mx}, we define the following $\chi^2$ function
\footnote{In the actual implementation, we assume the events to follow the
Poisson distribution. However, for illustrative purposes, it is sufficient
to consider the more compact approximative Gaussian expression.}
\begin{multline}
  \chi^{2} = \min_{\lambda} \sum^{\text{channel}}_{j} \sum_{i}^{\text{bin}}
    \frac{\left| N_{ij} \left( \lambda^{\text{true}}, \epsilon^{\text{true}} \right)
      - N_{ij} \left(\lambda, \epsilon = 0\right) \right|^{2}}
        {N_{ij} ( \lambda^{\text{true}}, \epsilon^{\text{true}})} \\
      + {\rm Priors},
\label{eq:chiSq}
\end{multline}
where $N_{ij}$ denotes the number of events in the $i$-th energy bin for the
oscillation channel $j$, the vector
$\lambda = ( \theta_{12}, \theta_{13}, \theta_{23}, \delta _{\rm CP}, \sdm, \ldm, a,
\vec{b} )$ contains the standard oscillation parameters, the Mikheyev-Smirnov-Wolfenstein
(MSW) potential $a$, and the systematical biases $\vec{b}$, and $\epsilon$ represents the
non-standard parameters. The index $j$ runs over the $\emu$ and $\mumu$ channels and over
the corresponding anti-neutrino processes. For the ``true'' parameters used to calculate
the simulated data, we adopt the following numerical values~\cite{Maltoni:2004ei}:
\begin{align*}
  \sin^{2} 2 \theta_{12}^{\text{true}} &= 0.83, \\
  \sin^{2} 2 \theta_{23}^{\text{true}} &= 1.0,  \\
  \sin^{2} 2 \theta_{13}^{\text{true}} &= 0.01, \\
  (\Delta m_{21}^{2})^{\text{true}}
    &= 8.2 \times 10^{-5} \text{ eV$^{2}$},     \\
  (\Delta m_{31}^{2})^{\text{true}}
    &= 2.5 \times 10^{-3} \text{ eV$^{2}$}.
\end{align*}
In the fit, we marginalize $\chi^2$ over all standard oscillation parameters
and over the systematical biases, but keep the non-standard parameters fixed
at 0. The prior terms implement external input from other experiments and have the
form $(x - x^{\rm true})^2 / \sigma_x^2$, where $x$ stands for any oscillation
parameter or systematical bias, and $\sigma_x$ is the corresponding externally given
uncertainty. We assume $\theta_{12}$ and $\sdm$ to be known
to within 10\% from solar and reactor experiments~\cite{Maltoni:2004ei},
and include a standard deviation of 5\% for the MSW potential $a$. All other
oscillations parameters are assumed to be unconstrained since the neutrino
factory itself has an excellent sensitivity to them. The systematical
uncertainties are summarized in Tab.~\ref{tab:sys}. 

\begin{table}
  \centering
  \begin{ruledtabular}
    \begin{tabular}{l|r|r}
                 & \bf $\nu_e$ appearance & \bf $\nu_\mu$ disappearance \\ \hline
      Signal     &     2.5\%              &   20\%                      \\ \hline
      Background &     2.5\%              &   20\%
    \end{tabular}
  \end{ruledtabular}
  \caption{Systematical flux normalization uncertainties in our neutrino
           factory setup~{\sf NuFact2}.}
  \label{tab:sys}
\end{table}

For compactness, our discussion will focus on non-standard interactions induced
by $\epsilon^{m}_{e\mu}$,  $\epsilon^{m}_{e\tau}$, and $\epsilon^{s}_{e\mu}$,
but of course, one could also derive similar results for the other possible
terms. In particular, NSIs in the detector can be expected to have similar
effects to those in the source.

Furthermore, we will always assume a normal mass hierarchy, both for the
simulated data and for the fit. The main influence of the inverted hierarchy
is to shift the atmospheric MSW resonance to the anti-neutrino channel, which
is in general less important for the overall sensitivity of the experiment because
of the smaller anti-neutrino cross section. However, one can easily see that
the discovery reach for non-standard interactions is robust with respect to the
presence or absence of the MSW resonance: The main effect of the resonance is
to enhance $|S_{\rm SO}|$ in Eq.~\eqref{eq:OscProb-NSI}. Therefore, if it is
effective, the signal term, which is proportional to $|S_{\rm SO}^* S_{\rm NSI}|$,
is enhanced. At the same time, however, also the standard oscillation background
proportional to $|S_{\rm SO}|^2$ will become larger. These two opposing effects
cancel each other, as can be seen from the $\chi^2$ expression~\eqref{eq:chiSq}:
If we assume $\lambda^{\text{true}} = \lambda$, the background terms
drop out in the numerator, but not in the denominator. Since, however, the numerator
contains an extra square, we obtain $\chi^2 \sim |S_{\rm SO}^* S_{\rm NSI}|^2
/ |S_{\rm SO}|^2$, i.e.\ the standard oscillation contributions cancel,
and the expression is unaffected by their MSW enhancement.
We have verified numerically, that our results would hardly be affected by using
the inverted hierarchy for the data and the fit, even if we included parameter
correlations and higher order terms.

\subsection{Effects proportional to $\epsilon^{m}_{e\mu}$}

\begin{figure*}[tbp]
  \begin{center}
    \includegraphics[width=6cm]{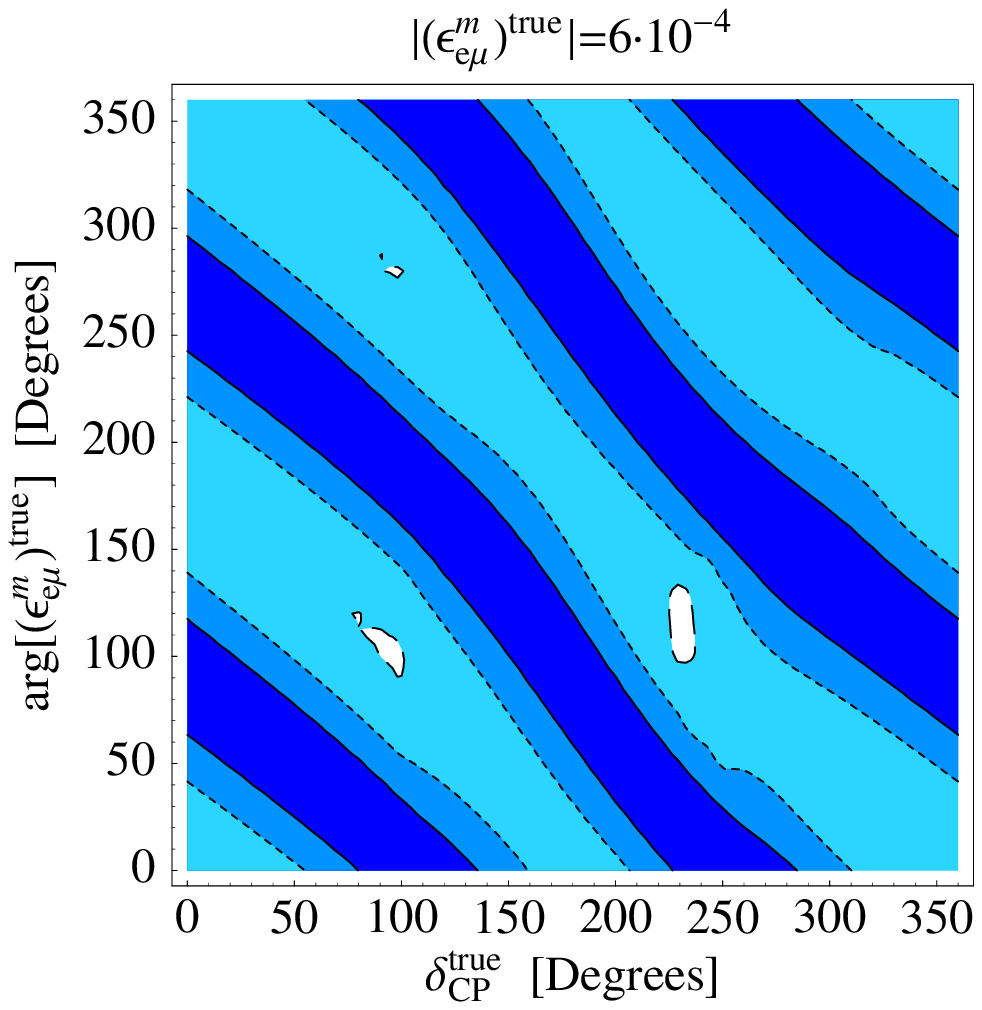} \\ \vspace{0.4 cm}
    \includegraphics[width=6cm]{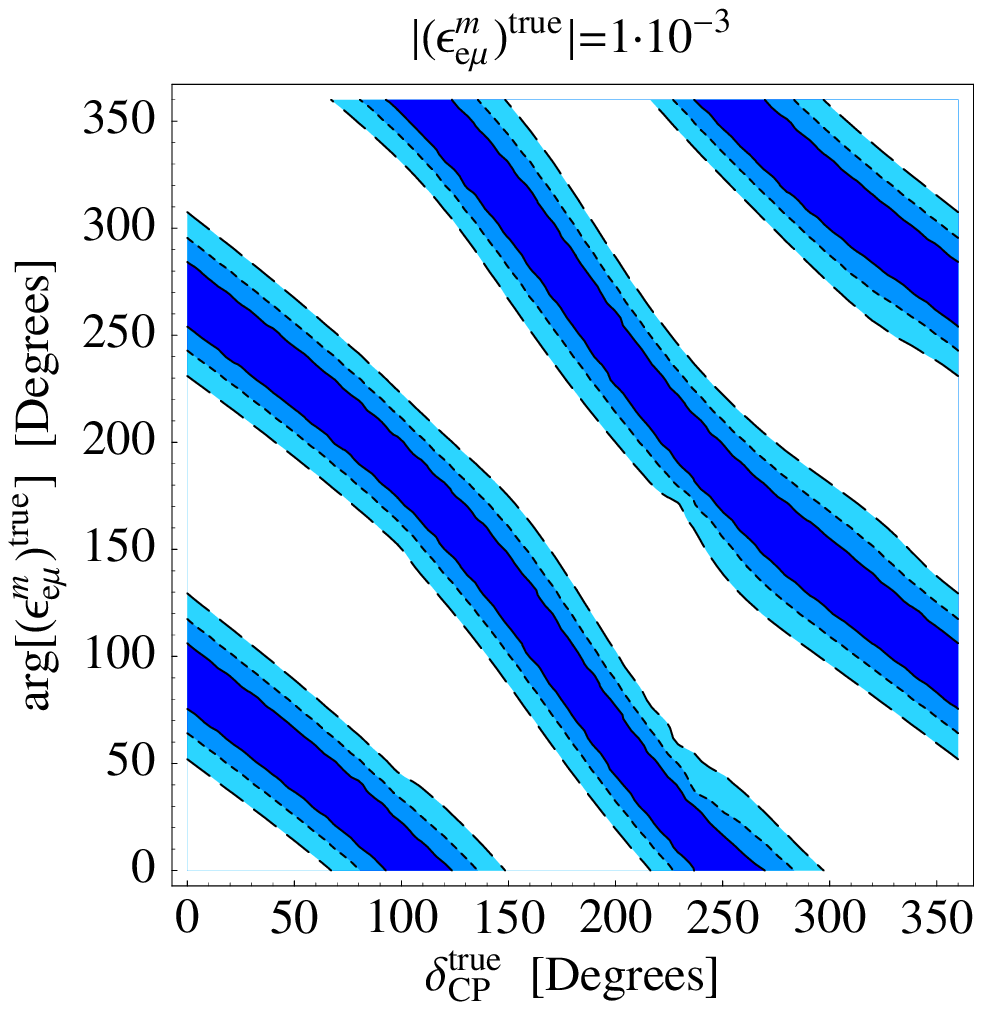}  \\ \vspace{0.4 cm}
    \includegraphics[width=6cm]{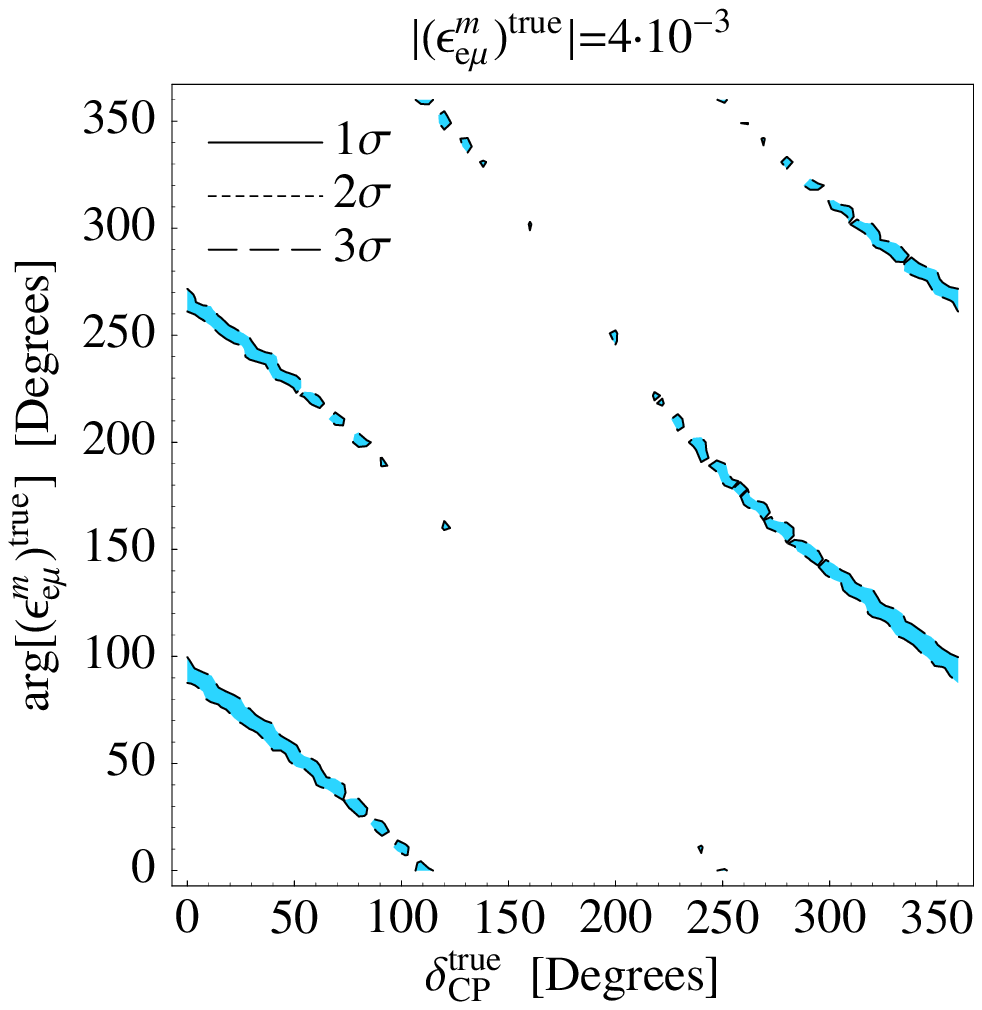}
  \end{center}
  \caption{Contour plots of the $\chi^{2}$ function defined in Eq.~\eqref{eq:chiSq}
    in the $\delta_{\rm CP}^{\rm true}$--$\arg[(\epsilon^{m}_{e\mu})^{\rm true}]$
    plane for $|(\epsilon^{m}_{e\mu})^{\text{true}}| = 6\times10^{-4}$ (top),
    $1\times 10^{-3}$ (center), and $4\times 10^{-3}$ (bottom). The value of
    $\sin^2 2\theta_{13}$ was taken to be $10^{-2}$ in all cases. The standard
    oscillation parameters and the matter potential were marginalized over.}
\label{Fig:island-plot}
\end{figure*}

We will first concentrate on non-standard effects proportional to
$\epsilon^{m}_{e\mu}$, and assume all other $\epsilon$ parameters to vanish.
Fig.~\ref{Fig:island-plot} shows the NSI discovery reach at $3\sigma$ as a
function of the true values of $\delta_{\rm CP}$ and
$\text{arg}[\epsilon^{m}_{e\mu}]$, and for three different
values of $|(\epsilon^{m}_{e\mu})^{\text{true}}|$. Since, according to
Eq.~\eqref{eq:chiSq}, $\text{arg}[\epsilon^{m}_{e\mu}]$
and $|\epsilon^{m}_{e\mu}|$ are fixed at zero in the fit,
while all other parameters are marginalized over, the contours are based
on the assumption of two degrees of freedom.\footnote{One might argue that the
leading term in the oscillation probability depends on the parameter combination
$\arg[\epsilon^{m}_{e\mu}] + \delta_{\rm CP}$ (see
Eq.~\eqref{eq:chi2} below), so it may be justified to use only 1~d.o.f. However,
since sub-leading contributions are not completely negligible, we take 2~d.o.f.\ to
be conservative.} If the parameters lie in the white regions of the plots,
non-standard interactions can be established at the $3\sigma$ level, while in
the dark areas, the sensitivity is less than $1\sigma$. It is obvious that for
larger $|(\epsilon^{m}_{e\mu})^{\text{true}}|$, the white regions of good
sensitivity become larger.

The characteristic band structure in Fig.~\ref{Fig:island-plot} reveals
that there are strong correlations between $\delta_{\rm CP}$ and
$\text{arg}[\epsilon^{m}_{e\mu}]$. To understand these correlations
analytically, we note that the leading NSI signal term is proportional to
$\cos(\arg[\epsilon^{m}_{e\mu}] + \delta_{\rm CP})$~\cite{Ota:2001pw}.
Therefore, the contribution of the $\emu$ channel to the $\chi^{2}$
function from Eq.~\eqref{eq:chiSq} becomes approximately
\begin{align}
  \chi^{2} \propto 
    \frac{\Big( \theta_{13}^{\rm true} \cdot |(\epsilon^{m}_{e\mu})^{\rm true}| \cdot
    \cos\big( \arg[(\epsilon^{m}_{e\mu})^{\rm true}] + \delta_{\rm CP}^{\rm true}\big) \Big)^{2}}
     {|(S_{\rm SO})_{\mu e}|^{2}}
  \label{eq:chi2}
\end{align}
and is thus expected to be roughly constant along the lines
of constant $\arg[(\epsilon^{m}_{e\mu})^{\rm true}] + \delta_{\rm CP}^{\rm true}$.
This behavior can be nicely seen in Fig.~\ref{Fig:island-plot}.

Note that correlations do not only limit the discovery reach for non-standard
interactions, but can also complicate the measurement of the standard
oscillation parameters~\cite{Huber:2002bi}.

Comparing the three plots in Fig.~\ref{Fig:island-plot}, we find that for
$|(\epsilon^{m}_{e\mu})^{\text{true}}| \sim 6 \cdot 10^{-4}$,
the first white islands appear, i.e.\ there are some parameter combinations
for which the non-standard effects can be discovered at $3\sigma$. For
$|(\epsilon^{m}_{e\mu})^{\text{true}}| \gtrsim 4 \cdot 10^{-3}$,
the $\chi^2$ values are above $3\sigma$ in the whole parameter space,
i.e.\ a 3$\sigma$ discovery is always possible, independent of
$\arg[(\epsilon^{m}_{e\mu})^{\rm true}]$ and  $\delta_{\rm CP}^{\rm true}$.
According to Eq.~\eqref{eq:epsilon-estimate}, these numbers translate into a
sensitivity to mass scales of up to $M_{\rm NSI} \sim 1 - 3$~TeV.

\begin{figure*}
  \begin{center}
    \includegraphics[width=12cm]{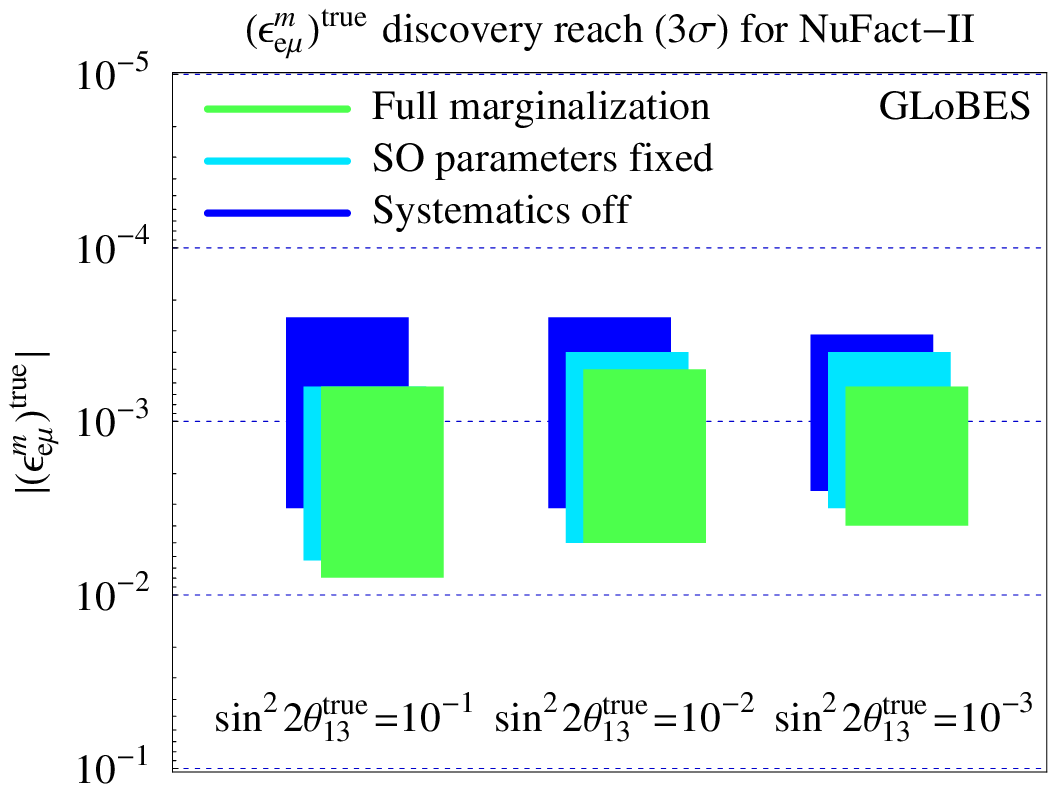}
  \end{center}
  \caption{Limitations to the discovery reach for $|\epsilon^{m}_{e\mu}|$
    arising from systematical errors and from parameter correlations.
    The top edges of the bars indicate the values of
    $|(\epsilon^{m}_{e\mu})^{\text{true}}|$ for which there exists some
    combination of $\arg[(\epsilon^{m}_{e\mu})^{\rm true}]$ and
    $\delta_{\rm CP}^{\rm true}$, which yields a sensitivity better than
    $3\sigma$; the bottom edges show how large the NSIs need to be in order
    to be detected at this confidence level for \emph{all} possible values of
    $\arg[(\epsilon^{m}_{e\mu})^{\rm true}]$ and  $\delta_{\rm CP}^{\rm true}$.
    The foremost (green) bars were obtained with the full analysis procedure
    discussed in the text, while for the intermediate (light blue) bars, the
    marginalization over standard oscillation parameters (including the MSW
    potential) was omitted, and for the hindmost (blue) bars, systematical
    errors were also switched off.}
\label{Fig:discovery-Eemu}
\end{figure*}

These values reappear as the top and bottom edges of the foremost (green)
bars in the middle part of Fig.~\ref{Fig:discovery-Eemu}. The bars stacked
below them (light blue) show how the discovery reach would improve if all
standard oscillation parameters and the MSW potential were known with
infinite precision, and the hindmost (blue) bars have been calculated
have been calculated under the additional assumption that systematical
errors are not present. The plot shows that the discovery reach depends
crucially on the true values of $\arg[\epsilon^{m}_{e\mu}]$ and
$\delta_{\rm CP}$, while systematical errors and the correlations with
the fit parameters have only moderate impact.

Comparing the results for different values of $\theta_{13}$, we find that the
achievable sensitivity for the most favorable combination of phase parameters
(top edges) remains roughly unchanged as $\theta_{13}$ decreases, while that
for the most problematic parameters (bottom edges) becomes slightly better. On
the one hand, smaller $\theta_{13}$ means smaller $(S_{\rm SO})_{\mu e}$, so
all terms in Eq.~\eqref{eq:OscProb-NSI} will decrease. On the other hand, the
standard oscillation background, which is given by $|(S_{\rm SO})_{\mu e}|^2$
and therefore proportional to $\theta_{13}^2$, will decrease faster than the
non-standard term, which is linear in $\theta_{13}$~\cite{Ota:2001pw}. This
makes it easier to disentangle signal and background, and especially when
correlations between the standard and non-standard parameters are taken into
account, this improved background suppression seems to overcompensate the
smaller signal.

Finally, let us compare the sensitivities predicted by our simulations
with existing bounds from charged lepton flavor violating
processes. In~\cite{Davidson:2003ha}, Davidson et al.\ constrain
$|\epsilon^m_{e\mu}|$ to be smaller than $\sim 8 \cdot 10^{-4}$ at the
90\% confidence level. Assuming the respective $\chi^2$ function to be
parabolic~\footnote{We are aware, that this extrapolation is problematic,
since realistic $\chi^2$ functions can be far from parabolic. However, it
should give a useful order of magnitude estimate.}, this translates into
a $3\sigma$ bound $\lesssim 1.4 \cdot 10^{-3}$. Comparing this number with
Fig.~\ref{Fig:discovery-Eemu}, we find that part of the parameter
space accessible by the neutrino factory is already ruled out, but, depending
on the phase correlation, our setup may still have a significant discovery
potential. This is particularly interesting if we note that
the present 90\% C.L. bound, $|\epsilon^m_{e\mu}| \lesssim 8 \cdot 10^{-4}$,
comes actually from three independent bounds on the coherent
forward scattering on up-quarks ($\lesssim 8 \cdot 10^{-4}$), down-quarks
($\lesssim 8 \cdot 10^{-4}$), and electrons ($\lesssim 5 \cdot 10^{-4}$). The
former two have been derived from $\mu \rightarrow e$ conversion in nuclei,
while the latter stems from the constraints on $\mu \rightarrow 3e$.
To improve the overall bound on $|\epsilon^m_{e\mu}|$ significantly,
all three components would have to be improved. Since no experiment
searching for $\mu \rightarrow 3e$ is being designed at the moment,
a neutrino factory seems to be the most realistic future option for
studying $|\epsilon^m_{e\mu}|$ in a model-independent way. Of course,
if the LHC should find evidence for one specific class of models, the
present bound might become much stronger already within the next few
years.

\subsection{Effects proportional to $\epsilon^{m}_{e\tau}$}
\label{Sec:epsilon-m-etau}

\begin{figure*}
  \begin{center}
    \includegraphics[width=12cm]{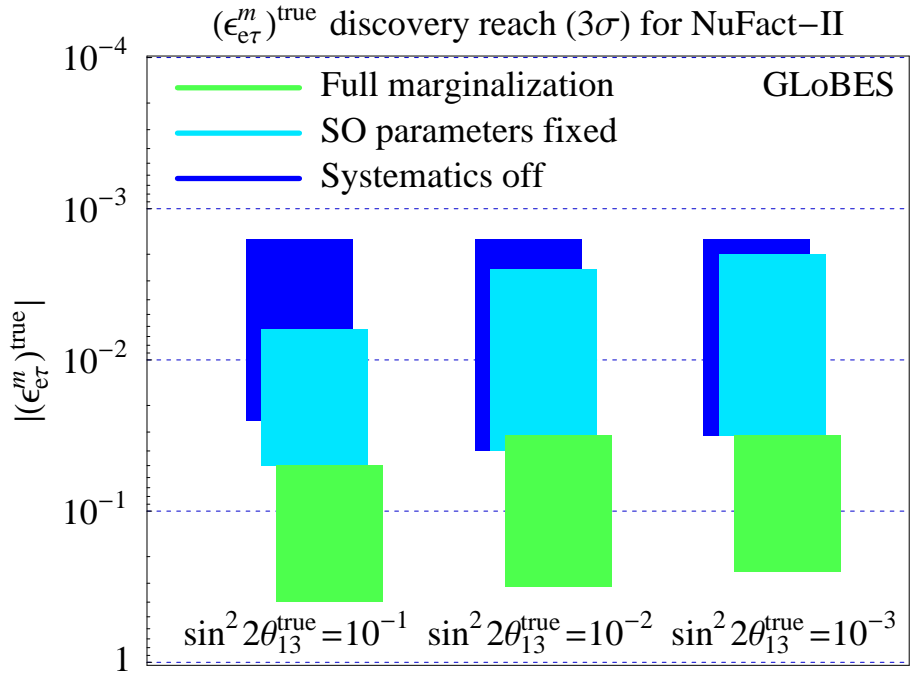}
  \end{center}
  \caption{Limitations to the discovery reach for $|\epsilon^{m}_{e\tau}|$
    arising from systematical errors and from parameter correlations.
    The color-coding is the same as in Fig.~\ref{Fig:discovery-Eemu}.}
\label{Fig:discovery-Eetau}
\end{figure*}

Let us now turn to non-standard effects proportional to
$\epsilon^{m}_{e\tau}$, which are introduced in analogy to
Eqs.~\eqref{eq:H_emu} -- \eqref{eq:H_emu_NSI}. It can be read off from
Fig.~\ref{Fig:discovery-Eetau}, that the sensitivity of a neutrino
factory to these effects is almost two orders of magnitude worse
than that to $\epsilon^{m}_{e\mu}$: Only for
$|\epsilon^{m}_{e\tau}| \gtrsim 3 \cdot 10^{-1}$,
discovery can be guaranteed. This can be understood from our
discussion in Sec.~\ref{Sec:Analytic}, which shows that the
energy dependence of standard and non-standard effects is the same,
so the effect of $\epsilon^{m}_{e\tau}$ can easily be absorbed
into $\lambda$. This also explains why fixing the standard oscillation
parameters improves the sensitivity by one order of magnitude.

As in the case of $|\epsilon^m_{e\mu}|$, the $\theta_{13}$ dependence in
Fig.~\ref{Fig:discovery-Eemu} is weak. Actually, it could be expected
to be larger here, because for the large values of $|\epsilon^{m}_{e\tau}|$
required for discovery, not only the interference term between standard
oscillations and NSI will contribute to the oscillation probability, but
also the pure NSI term proportional to $|\epsilon^{m}_{e\tau}|^2$. For
large $\theta_{13}$, both terms are comparable in magnitude, while for
small $\theta_{13}$, but still large $|\epsilon^{m}_{e\tau}|$, the pure
NSI term dominates. Therefore, in the latter case, the qualitative
behavior of the transition probability could be expected to be simpler
and thus more easily absorbed into $\delta_{\rm CP}$. However,
we see from Fig.~\ref{Fig:discovery-Eetau} that this effect is not
very pronounced.

The present bounds on $|\epsilon^{m}_{e\tau}|$ are of
$\mathcal{O}(1)$~\cite{Davidson:2003ha,Friedland:2005vy}, so the neutrino
factory could break new ground, independent of the phase correlations.
Note that in~\cite{Gago:2001xg}, the authors predict an even better performance
for the neutrino factory. However, they employ a completely different
experimental setup with a baseline of only 732~km, and use a different
analysis technique. Note also that it has been pointed out
in~\cite{Mangano:2001mj,Davidson:2003ha} that a measurement of the Weinberg
angle by neutrino scattering in the near detector of a neutrino factory could
improve the limit on $|\epsilon^{m}_{e\tau}|$ independently to well below 0.1.

\subsection{Effects proportional to $\epsilon^{s}_{e\mu}$}

\begin{figure*}
  \begin{center}
    \includegraphics[width=12cm]{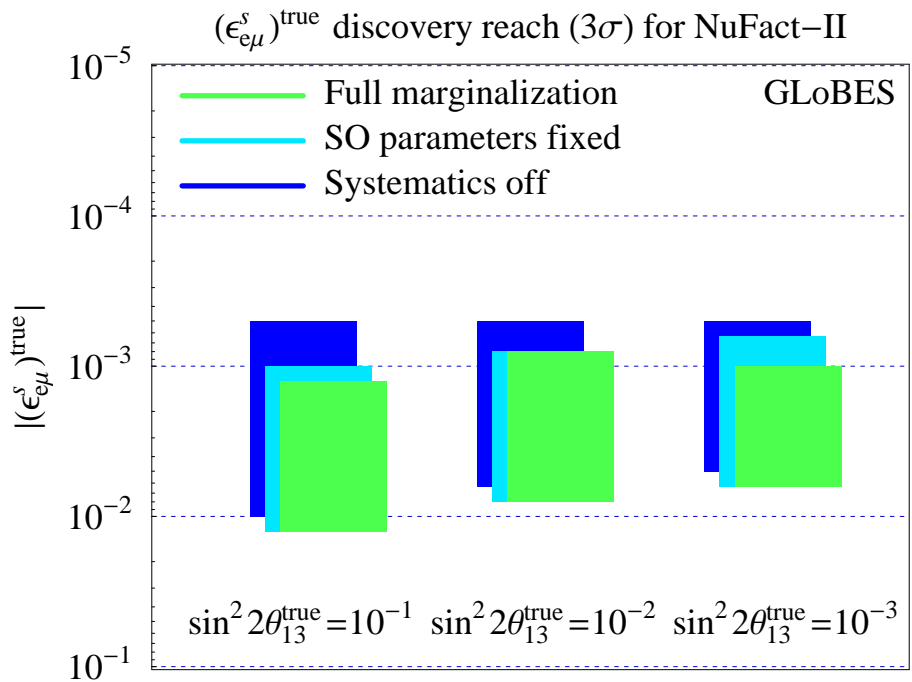}
  \end{center}
  \caption{Limitations to the discovery reach for $|\epsilon^{s}_{e\mu}|$
    arising from systematical errors and from parameter correlations.
    The color-coding is the same as in Fig.~\ref{Fig:discovery-Eemu}.}
\label{Fig:discovery-Semu}
\end{figure*}

If the non-standard interactions do not affect neutrino oscillations,
but rather the production process, we expect from Sec.~\ref{Sec:Analytic}
that the sensitivity will again be excellent because the standard
and non-standard terms have different energy dependence.
Indeed, Fig.~\ref{Fig:discovery-Semu} shows that effects with
$|\epsilon^{s}_{e\mu}| \sim 10^{-3}$ might be detected,
and that detection can be guaranteed for $|\epsilon^{s}_{e\mu}|
\gtrsim 10^{-2}$.

This discovery reach is at least one order of magnitude better than the
model-independent bound of $\mathcal{O}(10^{-1})$ coming from universality
considerations in lepton decays~\cite{Gonzalez-Garcia:2001mp}.
Let us, however, remark, that present \emph{model-dependent} bounds on
$\epsilon^{s}_{e\mu}$ can be already stronger than the sensitivity of the
neutrino factory.

\section{Conclusions}
\label{Sec:Conclusion}

We have investigated the prospects of a search for non-standard neutrino
interactions in a neutrino factory experiment. We have discussed several
different contributions that can arise in the effective Lagrangian, and
have pointed out that these can be distinguished from the standard
oscillations by their characteristic energy dependence. We have
performed careful numerical simulations of a neutrino factory experiment
to estimate its discovery potential for $\epsilon^{m}_{e\mu}$,
$\epsilon^{m}_{e\tau}$, and $\epsilon^{s}_{e\mu}$. It turned
out that there is a strong correlation between $\arg[\epsilon^{m}_{e\mu}]$
and $\delta_{\rm CP}$, so that the discovery reach for $\epsilon^{m}_{e\mu}$
depends strongly on the true values of these parameters: For certain
combinations, a discovery of the non-standard interactions is possible for
$|\epsilon^{m}_{e\mu}| < 10^{-3}$, while for less favorable scenarios,
$|\epsilon^{m}_{e\mu}| \sim 10^{-2}$ is required. Since the present
bounds on $|\epsilon^{m}_{e\mu}|$ are already of $\mathcal{O}(10^{-3})$, the
discovery potential will crucially depend on the phase correlations.
Vice-versa, however, a combination of neutrino factory results with
limits from other experiments might provide additional constraints on
the phases. The sensitivity to $|\epsilon^{m}_{e\tau}|$ is more than
one order of magnitude worse than that to $|\epsilon^{m}_{e\mu}|$ due
to the less favorable energy dependence of this effect. However, since
present bounds on $|\epsilon^{m}_{e\tau}|$ are very weak, a neutrino
factory could achieve a significant improvement here. Finally, the sensitivity
to $|\epsilon^{s}_{e\mu}|$ ranges between $10^{-3}$ and $10^{-2}$, which
is at at least one order of magnitude better than present model-independent bounds.
Thus, our simulations show that a neutrino factory is an excellent tool for
detecting new physics in the neutrino sector. However, reversing the
argument, this also means that possible non-standard interactions have to
be taken into account when analyzing the data of such an experiment.

\section*{Acknowledgments}

We would like to thank P.~Huber and W.~Winter for useful discussions.
This work was in part supported by the Transregio Sonderforschungsbereich TR27
``Neutrinos and Beyond'' der Deutschen Forschungsgemeinschaft.

\bibliography{./Discovery-NSI}
\bibliographystyle{apsrev}

\end{document}